\documentclass[showpacs,aps,graphicx,twocolumn]{revtex4}%and
\usepackage{graphicx}

\begin{document}

\title{Blind quantum machine learning}

\author{Yu-Bo Sheng,$^{1}$\footnote{shengyb@njupt.edu.cn} Lan Zhou,$^{2}$}
\address{$^1$Key Lab of Broadband Wireless Communication and Sensor Network
 Technology,
 Nanjing University of Posts and Telecommunications, Ministry of
 Education, Nanjing, 210003,
 China\\
 $^2$College of Mathematics \& Physics, Nanjing University of Posts and Telecommunications, Nanjing,
210003, China\\}

\date{\today }
\begin{abstract}
 Blind quantum machine learning (BQML)  enables a classical client with little quantum technology to delegate a remote quantum machine learning to the quantum server in such a approach that the privacy data is preserved. Here we propose the first BQML protocol that the client can classify two-dimensional vectors to different clusters, resorting to a remote small-scale photon quantum computation processor. During the protocol, the client is only required to rotate and measure the single qubit. The protocol is secure without leaking any relevant information to the Eve.  Any eavesdropper who attempts to intercept and disturb the learning process can be noticed. In principle, this protocol can be used to classify high dimensional vectors and may provide a new viewpoint and application for quantum machine learning.
\end{abstract}
 \maketitle

\section{Introduction} Quantum teleportation \cite{teleportation}, quantum key distribution (QKD) \cite{QKD,Ekert91}, quantum secure direct communication (QSDC) \cite{QSDC1,QSDC2} have been paid widely
attention for they can perform the absolute communication.  Quantum computing has also attracted much interest because of the discovery of applications that outperform the best-known classical counterparts. For example, Shor's algorithm  for integer factorization \cite{shor},  Grover's algorithm \cite{grover},  and the optimal Long's algorithm \cite{long,optimal} for unsorted database search, have shown the great computing power of quantum computers. The development of quantum algorithms is one of the most important areas of computer science. Recently, vast technological developments have been made for small-scale quantum computers in different quantum systems, such as ions \cite{ion}, photons \cite{photon}, superconduction \cite{superconduct}, and some other important quantum systems \cite{other1,other2}.

On the other hand, machine learning is a branch of artificial intelligence \cite{ml}. It learns from previous experience to optimize performance, which is widely used in computer sciences, robotics, bioinformatics, and financial analysis.  Generally, there are two kinds of machine learning. The one is supervised machine learning and the other is unsupervised machine learning.  By comparing the new Email with the old Email which has been labeled by human, the computer can successfully filter a spam after training. It is a supervised machine leaning. To recognize the object from a landscape background, that is, to classify the image pixels of the object from the background,  is an unsupervised machine leaning. Machine learning depends on the date base to perform the training. It is shown that the more data the computer can process, the more accurate of the model of machine learning is.
In machine learning, the most important algorithm in mathematical picture  can be described as follows: it is to evaluate the distance and inner product between two vectors. For  high-dimensional vectors, such task requires large time proportional to the size of the vectors. Therefore, the vector size will become a challenge for modern rapid growing big-data and the limitation of Moore's law in a classical computer.

In 2013, Lloyd \emph{et al.} showed that the quantum computer can be used to perform the machine learning \cite{lloyd}. Subsequently, there are several quantum machine learning (QML) protocols were proposed \cite{bang1,bang2,bang3}. In 2014, Bang \emph{et al.} proposed a method for quantum algorithm design assisted by machine learning \cite{bang1}. Yoo \emph{et al.} compared quantum and classical machines designed for learning a $N$-bit Boolean function \cite{bang2}. They showed that quantum superposition enabled quantum learning is faster than classical learning. Schuld \emph{et al.} provided an overview of existing ideas and approaches to QML \cite{qmlreview}.  Recently, Cai \emph{et al.} realized the first entanglement-based machine learning on a quantum computer \cite{pan}. Based on the linear optics, they both reported the approaches to implement the supervised and unsupervised machine learning.

In this paper, we will discuss another practical application for QML. In the QML, Alice (client) has some important and confidential data to implement the machine learning. However,  she does not have the ability to perform the QML. Fortunately, Alice has a rich and trusted friend named Bob (server), who has a quantum computer and can perform the QML. Can Alice ask Bob for help to perform the secure QML? Our protocol shows that it is possible for Alice to realize such QML. It is called the blind quantum machine learning (BQML).

This paper is organized as follows: In Sec. II, we first give the concept of BQML and describe a BQML protocol for classifying an arbitrary
two-dimensional vector to different clusters using a remote small-scale photonic quantum computer. In Sec. III, we will explain the secure of this BQML protocol. In Sec. IV, we will make a discussion and conclusion.

\section{Blind quantum machine learning} The concept of BQML can be detailed as follows: A client only can perform  single-qubit rotation and single-qubit measurement and dose not have sufficient quantum technology to delegate her QML. He/She asks a remote trusted server who has a fully fledged quantum power to perform the QML. During the process, any Eve who attempts to intercept and disturb the learning can be noticed.

We will describe a simple example to explain our BQML.
A key mathematical task of quantum machine learning algorithm is to assign a  new vector $\vec{u}$ to two different clusters $A$ and $B$ with one reference vector  $\vec{v_{A}}$ and $\vec{v_{B}}$ in each cluster \cite{lloyd, pan}. By comparing the distance $D_{A}=|\vec{u}-\vec{v_{A}}|$ and $D_{B}=|\vec{u}-\vec{v_{B}}|$, we can assign $\vec{u}$  to the cluster with smaller distance.
A quantum state has its natural advantage to  be used to represent a vector. Here we take the assigning  two-dimensional vector for example to explain the basic BMQL principle.
We let $\vec{u}=|u||u\rangle$, and $\vec{v}=|v||v\rangle$, respectively. $|u\rangle$ and $|v\rangle$ can be described as
\begin{eqnarray}
|u\rangle&=&\alpha|H\rangle+\beta|V\rangle,\nonumber\\
|v\rangle&=&\gamma|H\rangle+\delta|V\rangle,
\end{eqnarray}
with $|\alpha|^{2}+|\beta|^{2}=1$ and $|\gamma|^{2}+|\delta|^{2}=1$.
Here  $|H\rangle$ is the  horizonal polarization and $|V\rangle$ is the vertical  polarization of the photon, respectively. The distance between $\vec{u}$ and $\vec{v}$ can be described as
\begin{eqnarray}
D&=&|\vec{u}-\vec{v}|=\sqrt{|\vec{u}-\vec{v}|^{2}}\nonumber\\
&=&\sqrt{(|u|\langle u|-|v|\langle v|)(|u||u\rangle-|v||v\rangle)}\nonumber\\
&=&\sqrt{|u|^{2}+|v|^{2}-2|u||v|\langle u|v\rangle}.\label{distance}
\end{eqnarray}
From Eq. (\ref{distance}), the calculation of the distance can be converted to the calculation of the overlap of the quantum states $|u\rangle$ and $|v\rangle$.

\begin{figure}[!h]%[tpb]
\begin{center}
\includegraphics[width=8cm,angle=0]{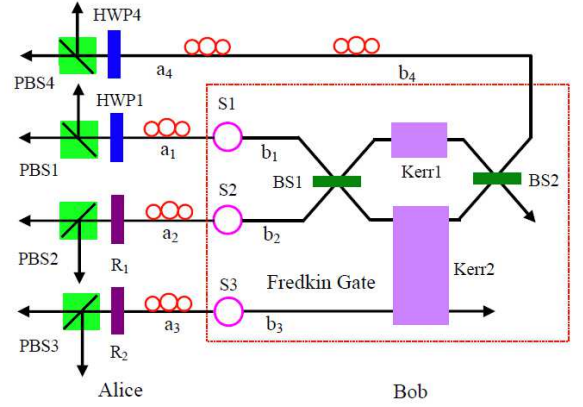}
\caption{Schematic of the principle of BQML. The Fredkin gate is constructed by the optical kerr nonlinearity \cite{kerr}. HWP is the half wave plate which can make $|H\rangle\rightarrow \frac{1}{\sqrt{2}}(|H\rangle+|V\rangle)$, and $|V\rangle\rightarrow \frac{1}{\sqrt{2}}(|H\rangle-|V\rangle)$. The PBS is
the polarization beam splitter. It can transmit the $|H\rangle$ polarized photon and reflect the $|V\rangle$ polarized photon.  $R$ is an arbitrary rotation for polarization photons. $S1$, $S2$ are the entanglement sources and $S3$ prepares the target single-photon state, respectively.}
\end{center}
\end{figure}

Our BQML protocol based on the optical system can be described as follows.

Step 1: As shown in Fig. 1, Bob first prepares three ordered $2N$  pair of state $|\phi^{+}\rangle$ using the entanglement sources $S1$,  $S2$, and $S3$, respectively. Here $|\phi^{+}\rangle$ is one of the Bell states of the form of
\begin{eqnarray}
|\phi^{+}\rangle=\frac{1}{\sqrt{2}}(|H\rangle|H\rangle+|V\rangle|V\rangle).
\end{eqnarray}
The $2N$ pairs of photons are divided into two groups. Each group contains $N$ pairs. The first group is named
checking group and the second group is named massage group.
The $2N$ pairs prepared by $S1$ is denoted as $|\phi^{+}_{1}\rangle_{a_{1}b_{1}}$, $|\phi^{+}_{2}\rangle_{a_{1}b_{1}}$, $\cdots$, $|\phi^{+}_{2N}\rangle_{a_{1}b_{1}}$. The $2N$ pairs prepared by $S2$ is denoted as $|\phi^{+}_{1}\rangle_{a_{2}b_{2}}$, $|\phi^{+}_{2}\rangle_{a_{2}b_{2}}$, $\cdots$, $|\phi^{+}_{2N}\rangle_{a_{2}b_{2}}$ and the  $2N$ pairs prepared by $S3$ is denoted as $|\phi^{+}_{1}\rangle_{a_{3}b_{3}}$, $|\phi^{+}_{2}\rangle_{a_{3}b_{3}}$, $\cdots$, $|\phi^{+}_{2N}\rangle_{a_{3}b_{3}}$, respectively. Subsequently, Bob sends one of the photons in each pair to Alice in each checking group in the channel $C_{a_{1}b_{1}}$,
$C_{a_{2}b_{2}}$, and $C_{a_{3}b_{3}}$, respectively. Here $a_{1}b_{1}$, $a_{2}b_{2}$ and $a_{3}b_{3}$ are the spatial modes as shown in Fig. 1. Subsequently, Alice and Bob check eavesdropping by the following procedure. a) Bob randomly measure his $N$ photons in the checking group in the basis $\{|H\rangle, |V\rangle\}$ and $\{|\pm\rangle=\frac{1}{\sqrt{2}}(|H\rangle\pm|V\rangle)\}$, respectively. b) Bob tells Alice that which basis he has used for each photon and all the measurement results. c) Alice uses the same measurement basis to measure all the $N$ photons and checks her results with Bob's. For  $|\phi^{+}\rangle$,
Alice and Bob always have the same measurement results in both  basis  $\{|H\rangle, |V\rangle\}$ and $|\pm\rangle$. If Bob obtains $|H\rangle$, Alice also obtains $|H\rangle$. If Bob obtains $|-\rangle$, Alice obtains $|-\rangle$ too. If some of the results of Alice and Bob are different, it means that eavesdroppers exists. In this way, they have to stop the BQML and check the quantum channel to eliminate eavesdropping.

Step 2: Bob sends another three sequences of $N$ pairs in three message groups to Alice, respectively. Certainly, After Alice receiving the photons, Alice and Bob can also randomly choose some photon pairs to perform the checking, which is the same as step 1 to ensure that there is no eavesdrop.

Step 3: Alice randomly measure the photons in channel $C_{a1b1}$ in the basis $\{|H\rangle, |V\rangle\}$ and $|\pm\rangle$, respectively. As shown in Fig. 1,
The HWP1 and PBS1 can be used to complete the task. Here
 HWP is the half-wave plate which can make $|H\rangle\rightarrow \frac{1}{\sqrt{2}}(|H\rangle+|V\rangle)$, and $|V\rangle\rightarrow \frac{1}{\sqrt{2}}(|H\rangle-|V\rangle)$. PBS is
the polarization beam splitter. It can transmit the $|H\rangle$ polarized photon and reflect the $|V\rangle$ polarized photon, respectively. After measurements,
Bob will obtain a random photon sequence in $C_{a1b1}$ such as $|H\rangle_{1}|-\rangle_{2}|V\rangle_{3}\cdots|+\rangle_{N}$. The polarization information of each photon is only known by Alice.

Step 4: Alice rotates her photons in channel $C_{a2b2}$. In Fig. 1, the operation $R_{1}$ can perform an arbitrary rotation as
$|H\rangle\rightarrow \alpha|H\rangle+\beta|V\rangle$ and $|V\rangle\rightarrow \beta|H\rangle-\alpha|V\rangle$. Subsequently, Alice lets her photon
pass through the PBS2 and measures it. Here  $|\alpha|^{2}+|\beta|^{2}=1$. If her measurement is $|H\rangle$, Bob's photon in $C_{a2b2}$ will become $|u\rangle=\alpha|H\rangle+\beta|V\rangle$.
On the other hand, if Alice's measurement result is $|V\rangle$, Bob's photon  in $C_{a2b2}$ will become $|u\rangle'=\beta|H\rangle-\alpha|V\rangle$.
In this way, Alice asks Bob to perform a bit-flip operation $\sigma_{x}=|H\rangle\langle V|+|V\rangle\langle H|$ and a phase-flip operation $\sigma_{x}=|H\rangle\langle H|-|V\rangle\langle V|$ to change $|u\rangle'$ to $|u\rangle$.  Similar to Step 4, Alice rotates her photon with $R_{2}$ in channel $C_{a3b3}$ and measures it to make the related photon in Bob collapse to $|v\rangle$.
 Then she asks  Bob to perform the Fredkin operation.

Step 5: Bob performs the Fredkin operation. The photons in spatial mode $b1$ is the control qubits and the photons in $b2$ and $b3$ are the target qubits, respectively. In an optical system, optical Kerr effect provides us a powerful tool to perform the quantum information processing. It can also be used to realize the quantum optical Fredkin gate  \cite{kerr}. As shown in Fig. 1,  if the control qubit is $|H\rangle$ or $|V\rangle$, the whole system can be evolved as
\begin{eqnarray}
|H\rangle|u\rangle|v\rangle\rightarrow|H\rangle|u\rangle|v\rangle,\nonumber\\
|V\rangle|u\rangle|v\rangle\rightarrow|V\rangle|v\rangle|u\rangle.\label{evove1}
\end{eqnarray}

On the other hand, if the control qubit is $|+\rangle$ or $|-\rangle$, the whole system can be written as
\begin{eqnarray}
|\pm\rangle|u\rangle|v\rangle
&\rightarrow&\frac{1}{\sqrt{2}}(|H\rangle|u\rangle|v\rangle\pm|V\rangle|v\rangle|u\rangle).\nonumber\\
&=&\frac{1}{\sqrt{2}}[|+\rangle(|u\rangle|v\rangle+|v\rangle|u\rangle)\nonumber\\
&\pm&|-\rangle(|u\rangle|v\rangle-|v\rangle|u\rangle)].\label{evove2}
\end{eqnarray}
After performing the Fredkin operation, the $N$ control qubits are sent back to Alice in channel $C_{a_{4}b_{4}}$. Fredkin operation does not change the polarization of control
qubit.  Alice know the exact polarization information of these $N$ photons. For the $k$th qubit, if it is $|H\rangle$ or $|V\rangle$, she measures it in
the basis $\{|H\rangle,|V\rangle\}$. On the other hand, if it is $|+\rangle$ or $|+\rangle$, she measures it in the basis $\{|\pm\rangle\}$.
The probability of obtaining $|+\rangle$ can be written as
\begin{eqnarray}
P_{+}=\frac{1+|\langle u|v\rangle|^{2}}{2}=\frac{1+|\alpha^{\ast}\gamma+\beta^{\ast}\delta|^{2}}{2}.
\end{eqnarray}
On the other hand, the probability of obtaining $|-\rangle$ can be written as
\begin{eqnarray}
P_{-}=\frac{1-|\langle u|v\rangle|^{2}}{2}=\frac{1-|\alpha^{\ast}\gamma+\beta^{\ast}\delta|^{2}}{2}.
\end{eqnarray}
Here $\alpha^{\ast}$ and $\beta^{\ast}$ are complex conjugate coefficients of $\alpha$ and $\beta$, respectively.
We can obtain that
\begin{eqnarray}
\langle u|v\rangle=\sqrt{1-2P_{-}}.
\end{eqnarray}
Compared with Eq. (\ref{distance}),
we can obtain
\begin{eqnarray}
D&=&\sqrt{|u|^{2}+|v|^{2}-2|u||v|\sqrt{1-2P_{-}}}.\label{distance2}
\end{eqnarray}
From Eq. (\ref{distance2}), similar to the approach of entanglement detection \cite{measureconcurrence,zhoupra}, the aim of calculation of $D$ is transformed to pick up the success probability of $P_{-}$. In a practical experiment, we should repeat this protocol many times to obtain a statistical accuracy simply by calculating the ratio between the detected photon number and the initial total photon number \cite{pan}.

\section{Security of blind quantum machine learning} Our BQML protocol is based on the Bell state $|\phi^{+}\rangle$. The proof for the security of our BQML protocol is based on the security for the first transmission of the photons prepared by $S1$ and $S2$ from Bob to Alice in channels $C_{a1b1}$, $C_{a2b2}$ and $C_{a3b3}$,
and the second transmission of the control qubit from Bob to Alice in channel $C_{a4b4}$.

In the first transmission, the security check in our protocol is similar to
the QSDC protocol \cite{QSDC2}. During the transmission, all the states
are the same Bell states $|\phi^{+}\rangle$. That is Bob does not encode any information to Alice. If Eve can capture one photon in each Bell state,
he gets no information. Once Alice and Bob share the same states $|\phi^{+}\rangle$ in both $C_{a1b1}$, $C_{a2b2}$ and $C_{a3b3}$,  by measuring the photons in the same basis
$\{|H\rangle,|V\rangle\}$ and $\{|\pm\rangle\}$ randomly, they always obtain the same measurement results. However, if Eve steals one photon and fakes another
photon to Alice, the faked photon does not entangle with the Bob's photon. By measuring the two photons, Alice and Bob will find that some of the photons
will induce the different measurement results, which shows that the Eve exists. Eve cannot elicit any information from the Bell states because there is no information encoded there. The information "comes into being" only after Alice perform measurements on her photons in   $C_{a1b1}$,  $C_{a2b2}$ and $C_{a3b3}$. However, after measurement, the photons in Bob's location collapse to the corresponded states, which shows that Eve cannot elicit any information during measurement.

The security of the second transmission of the control qubits from Bob to Alice is similar to the  QKD protocol \cite{QKD}. Note that Bob does not change the
polarization of the control qubits and he even does not know the polarization of these control qubits. Only Alice knows the exact information of these control
qubits. Therefore,  after Alice receiving these photons,  from Eq. (\ref{evove1}), if the initial control qubit is $|H\rangle$ or $|V\rangle$, she still
obtain $|H\rangle$ or $|V\rangle$ by measuring it in the basis $\{|H\rangle,|V\rangle\}$, respectively. Thus, if an unexpected measurement result occurs, for example, the initial state is $|H\rangle$ but the measurement result is $|V\rangle$ and vice versa,
  it means that the Eve has altered the photon.

  In BQML, the important and confidential data essential are $|u\rangle$ and $|v\rangle$.  After performing the BMQL, they are stayed at Bob's location.
  Therefore, the security of this BQML depends on the fact that  Bob is a trusted server. After Alice measuring the reference state in $C_{a4b_{4}}$, Bob essentially can obtain the information of $|u\rangle$ and $|v\rangle$ by measuring  a range of samples, for this BMQL is based on  a statistical accuracy simply by calculating the ratio between the detected photon number and the initial total photon number. In order to prevent Bob to intercept $|u\rangle$ and $|v\rangle$, Alice can asks Bob to send all the photons back in $C_{a4b4}$. For $|u\rangle$ and $|v\rangle$, the coefficients $\alpha$, $\beta$, $\gamma$
  and $\delta$ are unknown, Alice can notice whether Bob has measured these photons.
  
\section{Discussion and conclusion}

 So far, we have described our BQML protocol for classifying a two-dimensional vector. This protocol is also suitable for arbitrary
high-dimensional vector, based on the condition that  Bob has a powerful quantum computer processor to perform the Fredkin operation for high-dimensional quantum state and Alice can measure and operate arbitrary high-dimensional quantum state \cite{kerr}. In an optical system, the photons encoded in multiple degrees of freedom \cite{measureconcurrence,hyper} or the orbital angular momentum degree of freedom \cite{OAM1,OAM2} may be the good candidates to implement the high-dimensional vector. Recently an important work showed that  by exploiting the giant optical circular birefringence induced by quantum-dot spins in double-sided optical microcavities as a result of cavity quantum electrodynamics, it is possible to construct the controlled-CNOT gate to realize hyper-parallel photonic quantum computation  in both the polarization and  the spatial mode  degrees of freedom in a two-photon system simultaneously \cite{hypercnot}.
Certainly, in above description, we only discuss the BQML with ideal environment. Actually, for a practical application, we should consider the
environmental noise. It will degrade the maximally entangled state $|\phi^{+}\rangle$ to a mixed state. On the other hand, the distributed photons
should suffer from the photon loss. Fortunately, entanglement purification provided us the powerful tool to distill the low quality entangled state
to the high quality entangled state, which has been widely discussed in both theory and experiment \cite{shengpra1,shengpra2,shengsr,purification1,purification2,purification3,purification4,wangc}. Moreover, the quantum state amplification protocols have also been proposed to protect the single photon from loss \cite{amplication1,amplification2,amplification3,amplification4}. These progress showed that it is possible to realize the
BQML in future.

In conclusion, we have proposed the concept of BQML and described the  first BQML protocol on a photonic computation processor. Our protocol demonstrates that estimation of the distance between
vectors in machine learning can be naturally safely realized assisted with a remote quantum computer. This protocol combined with the previous researches of QML may provide a useful approach for dealing with the  "big data" and "cloud" computation.

\section*{ACKNOWLEDGEMENTS}
This work was supported by the National Natural Science Foundation
of China under Grant  Nos. 11474168 and 61401222, the Qing Lan Project in Jiangsu Province, and a Project
Funded by the Priority Academic Program Development of Jiangsu
Higher Education Institutions.\\

\end{document}